# Information and Protein Interfaces

William W. Chen[1], Paul J. Choi[2], Jason E. Donald[2], Eugene I. Shakhnovich[2]

[1]*Department of Biophysics, Harvard University, Cambridge, Massachusetts*
[2]*Department of Chemistry and Chemical Biology, Harvard University, Cambridge, Massachusetts*

*Correspondence to:

Eugene I. Shakhnovich

Department of Chemistry and Chemical Biology

Harvard University, Cambridge, MA 02138

e-mail: eugene@belok.harvard.edu

**ABSTRACT**

To confer high specificity and affinity in binding, contacts at interfaces between two interacting macromolecules are expected to exhibit pair preferences for types of atoms or residues. Here we quantify these preferences by measuring the mutual information of contacts for 895 protein-protein interfaces. The information content is significant and is highest at the atomic resolution. A simple phenomenological theory reveals a connection between information at interfaces and the free energy spectrum of association. The connection is presented in the form of a scaling relation between mutual information and the energy gap of the native bound state to off-target bound states. Measurement of information content in designed lattice interfaces show the predicted scaling behavior to the energy gap. Our theory also suggests that mutual information in contacts emerges by a selection mechanism, and that strong selection, or high conservation, of residues should lead to correspondingly high mutual information. Amino acids which contribute more heavily to information content are then expected to be more conserved. We verify this by showing a statistically significant correlation between the conservation of each of the twenty amino acids and their individual contribution to the information content at protein-protein interfaces

# INTRODUCTION

Interactions between macromolecules lead to the networks of genes and gene products that drive all biological processes. Recent yeast two-hybrid and mass spectrometry experiments permit the identification of these complexes on proteomic scales (1, 2). In turn, hypothetical protein-protein connectivity networks can be built from these data. In order to understand the physical rules that govern how macromolecules find and bind to each other *in vivo*, experimental x-ray structures of macromolecular complexes have been characterized by various metrics that have come to be regarded as important in biological complexes, such as buried surface area, planarity of the interface, hydrogen bond density, residue propensities and conservation (3-5). Another approach to understanding macromolecular interactions is via "docking", which involves computational modeling of the binding of two macromolecules by using different energy functions and models to reproduce the correct mode of binding (6-8).

However, while measurement of these putatively important metrics has given us better understanding of the problem, it is not a unifying approach since in order to measure any particularly important metric one has to know in advance what to measure. Further, it is unclear why particular computational models and algorithms fail and why others occasionally succeed in reproducing correctly bound conformations. Without *a priori* assumption of metrics or of energy functions and algorithms, a direct study of the sources of quantitative (or mathematical) information at molecular interfaces would deepen our understanding of the physical bases of how macromolecular complexes locate each other in cells.

We begin by considering that a macromolecular complex is expected to have a favorable free energy of binding, not only in isolation but also in the crowded environment of a cell. Because the environment of the cell is populated by many different species of macromolecules, two interacting macromolecules of a complex must compete against many possible promiscuous and off-target interfaces. These off-target interfaces can also be thought of as "decoy interfaces," as they must be recognized and avoided by the binding macromolecule. One strategy for the two macromolecules to discriminate each other against these decoy interfaces is to select strongly interacting residues specific and appropriate to each other's interfaces. On the level of contacts, types of residues or atoms of one interface should exhibit preferences to types of residues or atoms of the cognate interface. These preferences can be quantified in terms of the mutual information (MI) of contacts. Quantitatively, the MI of contacts is the amount of information gained about the pattern of atoms or residues on one interface, when the pattern of contacting residues or atoms on the cognate interface is revealed. The large and accumulating number of multimeric complexes in the Protein Data Bank (PDB) (9) contains enough statistics for estimating such information.

Earlier work by Cline et al (10) measured the MI of contacts in the interior of proteins, though these measurements were restricted to residue-residue contacts. In what follows, we measure and show substantial information in contacts at the atomic resolution for real macromolecular interfaces. Most satisfying would be a rigorous physico-chemical description of how the information relates to basic thermodynamic quantities and perhaps how such information arises. We therefore propose a simple physical model that relates the information in interfaces to the physics of binding, and test these relations with lattice interface models. An intriguing possibility is that the information content of atoms or residues leaves signals in the proteome or genome. Specifically, a prediction of the simple phenomenological theory is that conservation of a residue (or atom) is related to the amount of information it conveys. We investigate these biological implications by dissecting the information contribution residue by residue, and show that the information contribution and the conservation of a residue are linked in real protein-protein interfaces.

There is also information encoded in the shape of the macromolecular interface, as the patterns of contact formation depend upon interface shapes to be complementary and conducive to such contacts. However, one can imagine that information from shape and pattern are decoupled in the following way: the backbones of the two proteins in a complex can be held fixed, and the paired interface residues mutated. Such pairs can continue to interact because the interface shape remains largely the same. Indeed, it is in the spirit of this idea that two strategies have been delineated by investigators who study computational docking: shape complementarity and pattern matching (8). This approximate independence of shape and pattern of contacts suggests that measuring information in the pattern of contacts alone would provide a useful lower bound on the total information.

## RESULTS

### Mutual information in contacts

To measure the MI in interface contacts, a necessary first step is to divide residues or atoms into types, as per some typing scheme. We describe 4 residue-based typing schemes and 2 atom-based typing schemes below. The formal basis for calculating MI comes from the following equation (11):

$$(1) \quad I = \sum_{i=1, j=1}^{N} f(A_i, B_j) \log_2 \frac{f(A_i, B_j)}{f(A_i) f(B_j)}$$

The number of types is given by $N$. The frequency $f(A_i, B_j)$ is the joint probability of observing a contact between two residue or atom types $A_i$ and $B_j$. The frequencies $f(A_i)$ and $f(B_j)$ are the interface contact compositional frequencies- or how often a residue or atom is involved in a contact. These frequencies are estimated by counting the contacts and the identity of the atoms which make these contacts at biological interfaces. With the frequencies, Eqn. 1 can be used to estimate MI. Here, the MI of a contact is naturally interpreted as the reduction in uncertainty of the other residue or atom's type when the type of one of the species is known.

Mutual information was measured at both atomic and residue resolutions. For the residue contact measurements, four different intuitive typing schemes were used, similar to the schemes in work by Cline et al (10) which measured contact MI in protein interiors. Amongst these four are the hydrophobic and hydrophilic 2-residue typing scheme, and the standard 20-residue typing scheme. For the atomic contact measurements, two schemes were employed. In one, every distinct atom of a residue is given a type; this scheme leads to 83 total atom types and has been used for protein folding studies (12, 13). We excluded the 4 backbone atoms to focus on the more specific residue contributions which resulted in a typing scheme of 79 types. In the other atom typing scheme, certain atoms are grouped together based on chemical similarity. This typing scheme has been used successfully in protein fold decoy discrimination tests (14) and in ab initio folding simulations (15). We also excluded the 4 backbone atoms in this scheme, resulting in a reduced 23 atom typing scheme.

In general, measurement of MI will be inaccurate due to undersampling. To estimate the probability of an event $x$, one must calculate $p_{est}(x) = n_x / n_{TOT}$, where the estimated probability is given by $p_{est}$, $n_x$ is the total number of times event $x$ is observed, and $n_{TOT}$ is the total number of samples. The accuracy to which one can achieve in $p_{est}$ is limited by $n_{TOT}$, e.g.. if the true probability $p(x) = 0.1$, and $n_{TOT} = 5$ measurements are made, then the most likely estimate of $p_{est}$ would be $0$. Thus any measurement of MI will be a combination of a "true" MI with some component of inaccuracy arising from undersampling. To estimate the magnitude of the component arising from undersampling, control measurements were done on generated decoy complexes. A decoy complex is generated by shuffling the identities of residues or atoms while preserving the contact network. The MI of these shuffled complexes is then measured. This shuffled MI is subtracted from the measured MI of the PDB sets to give a corrected MI. Shuffling to estimate the non-informative component of MI was also used in the work by Cline et al (10). We verified that the procedure of subtracting shuffled MI from the measured MI gives a good, constant estimator of the unbiased MI that is independent of sample size, unlike the measured MI alone (see Supporting Information).

The MI per contact is shown in Table 1 for 895 protein-protein interfaces taken from previous work by Keskin et al (16). The set of 895 protein-protein interfaces were derived from the PDB, and are a combination of 109 structural classes, with each class populated by distantly related sequence members. Two-chain interfaces were parsed out from dimers, trimers, antibody-antigen complexes, etc. This ensures a good mix of different interfaces. A significant and large corrected MI appears for the atomic typing schemes. While the residue typing schemes show non-zero statistically significant corrected MI per contact, the average number of residue-residue contacts at an interface for this set of PDB structures is 29, compared to 174 for atom-atom contacts. Therefore, not only is the information per contact lower, but the total information content is substantially lower in residue-residue contacts. For these reasons, it seems that the information of the entire interface is primarily driven by the atom-atom contacts.

**Selectivity, Gap and Information in Protein Interfaces: a Simple Statistical Mechanical Theory**

To gain insight into the biophysical origins of the measured MI, we introduce a simple phenomenological theory. Fink and Ball have considered the role of information and stability for folded proteins (17). We apply a similar formalism to estimate the "designed" or native state energy of the complex. We also consider the energies of an interface binding to randomly generated decoy interfaces, which represent the various promiscuous interactions in the cell environment.

We describe an interface as having $n$ sites on which chemical species are situated. The total number of possible chemical species is given by the typing scheme size $A$. For greatest generality, the chemical species may be atoms or residues. The identities of the atom or residue types at each site of the probe interface is given by $\{s_i\}$, and at each site of the target interface by $\{s_i^0\}$. The following coarse-grained Hamiltonian is adopted for describing the interaction energy of paired interfaces.

(2) $$E = \sum_{i=1, j=1}^{n} u_{s_i, s_j^0}$$

The term $u_{s_i, s_j^0}$ gives the pairwise interaction energy of contact between two chemical species types, one from each interface of the complex. n is total number of contacts at the interface. Rather than assume an explicit pairwise energy matrix between the different types, we assume only that the distribution of these interaction energies is Gaussian. The composition of the interface is assumed to be unbiased, so that this Gaussian assumption is reasonable.

(3) $$p(u_{s_i, s_j}) \propto \exp\left(-\frac{(u_{s_i, s_j} - u_{AVG})^2}{2\sigma^2}\right)$$

The mean and variance of this distribution are given by $u_{AVG}$ and $\sigma$ respectively. Because the total interaction energy is a sum of n independent pairwise interface contact energies, the distribution of decoy energies will also be Gaussian. Here, by decoy interfaces, we mean not only the incorrect protein targets, but also different orientations and surfaces of these species. That is, if two interfaces in some bound state are rotated relative to each other at the interface, then this will approximately generate a series of decoy states since the contacts each interface sees will change. An important parameter in this Gaussian distribution is the value $E_C$, which is an estimate of the lowest energy in this set of decoys, and can also be interpreted as the part of the Gaussian spectrum of interaction energies where states became "scarce". The parameter $E_C$ is given by the following expression from extreme value statistics (18, 19):

$$(4) \quad E_C = E_{AVG} - \sqrt{n\sigma^2 2\log\left(\frac{M}{\sqrt{2\pi}}\right)}$$

where M is the total number of decoys and $E_{AVG} = nu_{AVG}$ is just the average energy of the interface when paired to another interface selected at random.

Similarly, given a probe site and residue or atom type on that site, we can also estimate the energy of interaction with the native residue or atom type on the target site, which presumably has undergone some optimization or selection. To model the generation of an optimized target site, we begin by defining the selection strength $S$ of the target site as follows: a residue or atom type is selected $S$ times, randomly and with replacement from the set of $A$ possible types (in the case of atom types, this description of selection is only approximate, since atoms cannot be selected without also selecting the parent amino acid). Typically, after $S$ rounds of selection, not all possible $A$ types will have been encountered at a site, and some types will even have been encountered multiple times. The size of the set of amino acids that are encountered is given by $A_{eff}$, the effective size of the alphabet that a site sees under selection pressure $S$. Qualitatively, it is easy to see that a low selection strength $(S < A)$ will only draw upon a small sample of chemical species such that $A_{eff} < A$;

conversely, when $S \gg A$, then $A_{\it eff}$ will more likely be close to or equal to $A$. The detailed dependence of $A_{\it eff}$ on $S$ is not vital. Out of the $A_{\it eff}$ types encountered during selection at the target site, the type with the lowest pairwise energy to the type sitting on the probe site is kept. Thus the selection strength dictates how extensively a target site scans the space of available atom or residue types for optimization, and may be thought of as being crudely akin to the evolutionary pressure at a binding interface. Sites under greater selective pressure (higher $S$) sample a greater space of residues or atoms (and thus leading to higher $A_{\it eff}$) for a favorable interaction. Sites under little pressure will most assuredly pick a random interacting partner on the target interface with little contribution to the binding energy. We use extreme value statistics to approximate the energy obtained from this selection procedure. We assume that this energy is the lowest of $A_{\it eff}$ randomly selected values from the distribution in Eqn. 3:

$$(5) \quad E_N = E_{AVG} - \sum_{i=1}^{n} \sigma \sqrt{2 \log \left( \frac{A_{\it eff}^i(S)}{\sqrt{2\pi}} \right)}$$

Here summation is taken over all sites and $A_{\it eff}^i$ is site-specific effective size of the set of encountered aminoacid or atom types.

We can define an energy gap ratio $\varepsilon$, as in the work by Fink and Ball (17), to characterize thermodynamic properties of the probe interface and the multitude of decoy interfaces. There is an energy gap between the energy of the "best", lowest energy random decoy and the average energy in Eqn 4. There is also an energy gap between the energy of the native bound state and the average energy in Eqn 5. The energy gap ratio $\varepsilon$ is defined as a ratio of the two:

$$(6) \quad \varepsilon = \frac{E_N - E_{AVG}}{E_C - E_{AVG}}$$

The three quantities on the RHS are shown schematically in Fig 1. A higher ratio $\varepsilon$ means the natively bound interfaces are far separated from the decoy states. Using the relations in equation (4) and (5), the energy gap between the native bound state and the spectrum of promiscuously bound decoy states can be calculated:

$$(7) \quad \varepsilon^2 = \left(\frac{E_N - E_{AVG}}{E_C - E_{AVG}}\right)^2 = \frac{n \log\left(\frac{A_{eff}}{\sqrt{2\pi}}\right)}{\log\left(\frac{M}{\sqrt{2\pi}}\right)} = \frac{MI}{\log_2\left(\frac{M}{\sqrt{2\pi}}\right)}$$

The denominator of Eqn. 7 is simply the information required in selecting out one interface of $M$ decoys. The numerator is the information arising from selecting one type out of $A_{eff}$ possible ones during the interface selection process. Because the only source of information is in the specificity of selection of a target residue or atom type with respect to some probe residue or atom type, we interpret the mutual information to be the same as the numerator. Eqn 7. shows that there is a close connection between pushing the energy gap of the natively bound state below the energies of the promiscuous states, and the information required to maintain this tight and specific binding due to selection of atoms and residues on the interface..

We tested Eqn. 7 with a simple lattice interface model. A schematic of two lattice interfaces is shown in Fig 2. Lattice interfaces were composed of $n = 25$ sites. A general lattice species typing scheme of $A = 50$ types was used, along with a $50 \times 50$ pairwise potential with Gaussian random entries. Energies between two interfaces are calculated by aligning two interfaces and summing over the 25 corresponding sites as in Eqn 2. Sets of 1,000 probe interfaces with random lattice types (chosen from the 50) at each site were generated. For each probe interface, a target interface was designed. Because Eqn. 7 was motivated with the help of the selection strength parameter $S$, which is appealing physically and biologically, we can easily introduce $S$ into the lattice design procedure to design interfaces with different energy gaps. At each site, residue types were selected randomly and with replacement $S$ times, the one with best energy to the residue type on the corresponding probe site was retained (see Methods for details of lattice interface design). Done over all sites, this procedure generates a designed interface. In the end, a set of 1,000 probe and matching target interfaces was made. Each set has a characteristic energy gap due to the selection strength $S$ used. We treated the interfaces in the same manner as the PDB interfaces by calculating MI of the set with tabulated frequencies of contacts, via Eqn 1. Importantly, the energy gaps for the designed interfaces were also estimated by recording the average and best energies to the probe interface against random decoy interfaces. We used 10,000 random decoys to estimate $E_{AVG}$ and $E_C$. In Fig. 3, the MI versus the energy gap is shown for each set of thousand designed interfaces. Each point represents the average energy gap and MI for a set of designed interfaces at one value of $S$. A clear quadratic dependence between the MI and the average energy gap is observed. As the selection strength $S$ is tuned lower, the energy gap and information at the interfaces diminishes in the predicted fashion.

We remark here that the selection strength $S$ is closely related to the frequently measured quantity conservation. Sites with high selection strength $S$ will tend to have higher $A_{eff}$, as these sites have sampled the total possible $A$ types more thoroughly. Because these sites have a larger effective alphabet, it is more likely that they have hit upon the "optimal" partner species. In repeated design attempts, the sites with high $S$, from one design run to another, will then tend to be occupied by just a few "optimal" species. One would consequently observe good conservation at these sites in different design runs. According to Eqn. 7, sites with high $S$ will also have more information. Given this, we expect that the same atom or residue types which give high information content will also be the same ones which will be highly conserved in homologous interfaces.

**Sequence Conservation and Information in Interface Residues**

Our phenomenological theory suggests that conservation and information should be correlated to each other. Because amino acid conservation can be measured for protein-protein interfaces, we turn our focus to information content and conservation at the residue level. We begin by calculating the information contribution from one residue due to its constituent atoms. This can be done by looking at its component in the MI (Eqn. 1).

$$(8) \quad I = \sum_{i=1, j=1}^{N} f(B_j) \frac{f(A_i, B_j)}{f(B_j)} \log_2 \frac{f(A_i, B_j)}{f(A_i) f(B_j)} = \sum_{j=1}^{N} f(B_j) \left[ \sum_{i=1}^{N} f(A_i | B_j) \log_2 \frac{f(A_i | B_j)}{f(A_i)} \right]$$

Here the conditional probability $f(A_i | B_j)$ is equal to $f(A_i, B_j) / f(B_j)$. On the right hand side of Eqn. 8, the term in square brackets is the information "gain" due to a single atom type $j$. Intuitively, this can be seen as a comparison between the conditional composition of types $A_i$, and the background (or natural) composition of $A_i$. The physical meaning is this: when the composition of chemical types that are in contact with type $j$ is the same as the background composition of chemical types, then there is no additional knowledge obtained from knowing type $j$ is in the contact. However, when the two compositions are different, then the information gain is always greater than 0. The information gain is also formally known as the Kullback-Leibler distance between two distributions. $I_j^{GAIN}$ in Eqn. 9 gives the information contribution or information "gain" of atom type $j$.

$$(9) \quad I_j^{GAIN} = \sum_{i=1}^{N} f(A_i|B_j) \log_2 \frac{f(A_i|B_j)}{f(A_i)}$$

We calculated the information gain of every residue at atomic resolution with the expression in Eqn. 9 (see Methods). Table 2 shows the information gain for each of the 20 amino acids. Most interesting is that the charged residues aspartate, glutamate, lysine and arginine all show up in the top 6. These are expected, as charged residues are very often complementary at interfaces, thus being highly informative in the sense that one would expect a contact partner to be a residue of the opposite charge. Two of the aromatic amino acids, phenylalanine and tryptophan (but not tyrosine) are also in the top 6.

To compare information with conservation of amino acid types, we turn to a recent work by Keskin et al. In that work, structural alignments were performed on 3,799 representative interfaces, including low-homology structures to each of these representatives (20). The alignments allow one to calculate conservation statistics for each of the twenty amino acids. We ranked the 18 studied amino acids by both the calculated information gain, and by conservation. The ranking of the 18 amino acids according to conservation is the following: F, W, Y, R, L, D, P, I, H, E, M, Q, N, S, V, T, K, G. Interestingly, comparison of the Keskin et al ranking of amino acids and our information ranking (Table 2.) shows that the first few amino acids are very good matches. In particular, the charge pairs arginine and aspartate, and the bulky residues phenylalanine and tryptophan emerge as highly ranked residues in both conservation and information. In more systematic analysis, we performed a Spearman ranked correlation test between the information ranked and the conservation ranked amino acids. The results are shown in Fig 4. What is seen is a statistically significant correlation of 0.63 ($p < 0.003$) between the two. It is important to consider that conservation is probably not dictated solely by the information requirements between two interfaces, but perhaps also by constraints on shape of the residue or compositional requirements of surfaces. Despite this, the fact that there is a statistically significant correlation in the rankings between information and conservation is an encouraging confirmation of a prediction of our theory, and specifically points out that information may play some direct role in determining why certain residues are conserved.

## Discussion

The results of this study show statistically significant information exists at the atomic level, whereas at the residue level, the information is substantially lower. An apparent reason for that is that the residue representation is too coarse. While real "informative" contacts are present in native complexes, many spurious contacts will also be generated with a residue representation. It should be noted here that while evolutionary selection at protein interfaces occurs at the level of aminoacid residues there is still significant flexibility of choice of atomic contacts due t possibility of rotational isomerization of side chains on the interfaces. Such flexibility (almost) amounts to pissbility of independent selction of atomic contacts – through choic of both aminoacid chemical identity and its rotational isomerization state.

To explain the physical origins of the information, we used a simple phenomenological theory to relate this information to features of the binding energy spectrum. The simple theory of interfaces led to a consideration of the competition between the native target and the multitude of promiscuous off-target interfaces. The information was shown to be intimately linked to the ability of a probe interface in being able to successfully overcome the population of promiscuous off-target interfaces. Analysis of simple interface models confirm the predictions.

Importantly, our phenomenological theory is highly suggestive of the fact that conservation and the amount of information gain of a residue are linked. This was confirmed by a statistically significant positive correlation between ranking of conservation levels of each amino acid and the ranking of information contribution of amino acids. These results also raise the possibility that more genomic or proteomic signatures of information remain to be uncovered, beyond that of simple conservation of amino acids.

We make here a distinction between the different ways of using information to look at complexes. On the system level, the fact that an interface is able to pick out its cognate partner out of $M$ possible interfaces (drawn from different orientations of different targets) means that there are at least $\log_2 M$ bits encoded in this interface, i.e. once given the probe interface, we can reduce the uncertainty in the number of possible cognate interfaces from $M$ down to one.

This information, however, must arise from some microscopic source. An interface is microscopically described, roughly, by its size, shape, and pattern of atoms or residues. Therefore, we may view the information content in another way by considering the pattern of contacts that appear at the interface. The appropriate way to interpret the information due to the patterns is to ask, given the pattern of atoms or residues at the probe interface, what is the reduction in uncertainty of the pattern of contacts in the cognate interface? This important connection between the MI of contacts and the information content at interfaces means that the contact MI is both a microscopic basis for information content, and also a practical route whereby one can make measurements of information in macromolecular structures.

In our discussion, we have also neglected the role of topography or shape at interfaces. For protein-protein interactions, it is generally agreed that the two surfaces at the interface are fairly complementary in shape. The degree of complementarity, if quantified, can give a route to measuring the information due to shape alone. In the formulation of information theory then, one would ask, given the shape of one surface of a complex, how much uncertainty is reduced in knowledge of the shape of the other surface? If interfaces are typically well-fitted, then knowing the shape of one partner conveys tremendous information about the shape of the other partner, thereby producing a large reduction in uncertainty. Correspondingly, if interfaces are usually poorly fitted, knowledge of the shape of one partner reveals little about the other, thereby producing a small reduction in uncertainty. Another detail we have neglected is the dynamics of side-chains (or bases) at the interface. The overall effect of dynamic side-chains increases the number of decoy interfaces, as changing conformations will make the interface into slightly different targets. Naturally, this makes it more difficult for the probe molecule to find its target.

The estimates of MI per contact can be interpreted physically, despite being small fractional values. For example, in the protein-protein set, the corrected MI is 0.16 bits per atomic contact. One way to interpret this is that when one is given the identity of one atom in a contact, the uncertainty of the other atom is reduced effectively from 79 types down to 71 types ($\log_2 79 - \log_2 71 \approx 0.16$ bits).

We can speculate on the total information contained at typical macromolecular interfaces and its implications for biological function. For protein-protein interfaces, the measurement of 0.16 bits/contact at an average of 174 contacts/interface gives about 28 bits per interface. Therefore, assuming 28 bits at an interface and following Eqn. 6, we conclude that to achieve an energy gap ratio of at least $\varepsilon \approx 1$, the number of decoy interfaces can be at most $1 \times 10^8$. That is, if the requisite energy gap is near 1, a macromolecule can still find its target in a space of a hundred million non-target interfaces. This fact suggests at least, that with this type of atomic contact model, there should be some energy function and computational algorithm that exists which can guide one member of a complex to find the correct target site amongst $1 \times 10^8$ decoy sites. The caveat of course, is that the one must keep the backbone and side-chains of the macromolecule interfaces static as they are in the bound conformation, and that the requisite energy gap should be near 1. As for biological constraints on the magnitude of energy gaps of macromolecular interactions, these most likely depend upon details such as biological function. For fast enzymatic reactions between two proteins, a low energy gap may suffice since only a transient interaction is required to fulfill biological function. In a signaling pathway with a sustained response or binding, a large energy gap may be required. In reality, the actual binding of two interfaces would also depend upon the concentrations of the cognate-target pair and decoy interfaces as in a biological context. A full consideration of how concentration affects the requisite energy gap and information content is interesting and would require more detailed calculation.

In general it is difficult to know, *a priori*, how much detail to incorporate into simulations of macromolecule complexes docking against each other. The measurement of large MI only at the atomic level suggests that using MI can help guide or modulate the level of detail in model building before conducting expensive simulations. This explains why some recent successes have been achieved with atomic models (7), and that perhaps future docking approaches should also concentrate on atomic representations to maximize their chances for success. The increase in detail from residues to atoms is accompanied by an increase in the information available at the interface, which in turn should lead to a larger energy gap between the native state and incorrectly complexed states. When using an energy function to score and recognize the native binding state, a higher energy gap is more likely to lead to greater success in finding this state through simulations, as is the case for protein folding or fold recognition (19, 21).

## Methods

In the protein-protein measurements, MI was measured for 895 protein-protein interfaces used in previous studies (16, 20). (See Supporting Information for a list of PDB file names)  The interfaces were extracted from the PDB (PDB names are listed in Supplemental Materials) and represent a diverse set of structures and sequences, after having being filtered for both sequence and structural redundancy.  The information we observe, should therefore be quite general and not biased by peculiarities in the PDB. For the residue models, a cut-off of 8 Å between C-alpha atoms was used to define a contact. For the atom models, the cut-off was atom-specific, depending upon an effective van der Waals radius as used in previous work (22). This contact cut-off between two atoms is obtained by multiplying the sum of van der Waals radii of the two atoms by a constant "interaction" factor of 1.4 (22). Shuffled controls of MI were generated by repeatedly swapping atom identities at one interface for all complexes while maintaining the contact network.

Computationally generated interfaces were made using a lattice model. A lattice species type set of $A = 50$ lattice types is defined and given physical meaning with an explicit $50 \times 50$ pairwise potential. The lattice types represent, for generality, atoms or chemical species. The potential is composed of random Gaussian entries with mean 0 and standard deviation 1. The interface size was set to $n = 25$ sites. 1,000 different probe interfaces were generated by randomly assigning lattice types to the 25 sites. Target interfaces were designed at different selection strength $S$, in the following way: first, for each atom or residue type in the probe, $S$ types are selected at random with replacement; second, the type with the lowest energy to the probe is affixed to the corresponding position in the target interface. After the target interface is built up, the probe and target binding energy $E_N$ is calculated from Eqn. 2. For each of these probe interfaces, 10,000 random interfaces were also generated for decoys. A random interface was generated by placing randomly on each of the 25 lattice type sites one of the 50 lattice species type. Energies were calculated as in Eqn 2. between the random interface and the probe interface. The most favorable interaction energy out of these 10,000 decoys was taken to be $E_C$, and the average energy of these decoys was recorded as $E_{AVG}$. The energy gap was measured with these estimates. Mutual information was measured for contacts seen in these designed complexes in the same way MI was measured for protein-protein complexes, by tabulating single and pair frequencies of contacts and using Eqn. 1. The energy gap was averaged over the 1,000 probe-target pairs.

Information gain (or Kullback-Leibler distance to the background distribution) for each residue was calculated on the atomic level. The 79 atom typing scheme was used, as these distinguish the amino acids more readily than the reduced 23 atom typing scheme. In the 23 atom typing scheme, some atom types are shared between amino acids. First, the information gain due to each one of the 79 atom types was calculated from Eqn. 9, using the same single and pair contact frequencies used to calculate MI in Eqn. 1. Second, for each amino acid, the information gain of each atom is summed up to give a total information contribution. For example, arginine contains 8 atoms, which are the C-$\beta$, C-$\gamma$, C-$\delta$, N-$\epsilon$, C-$\zeta$, and two N-$\eta$ atoms. These are respectively typed 2, 3, 4, 5, 6, and 7. The information gain for arginine would be defined as the sum of the information gain of each of the 8 atoms.

## Acknowledgements

We would like to thank Drs. Igor Berezovsky, Jae Shick Yang and Konstantin Zeldovich for discussion and reading the manuscript. This work was supported by the NIH.

**Table Legends**

**Table 1.** The measured MI per contact of PDB structures of protein-protein interfaces. 895 interfaces from Keskin et al (16) were used. Corrected MI is the difference of the MI between PDB and the decoy complexes. [1] Hydrophobic / hydrophilic typing [2] Positive / Negative / Uncharged typing [3] Positive / Negative / Hydrophobic / Hydrophilic / Other typing [4] 20 residue typing [5] 23 chemically similar protein atoms typing [6] 79 distinct protein atoms typing.

**Table 2.** The information gain due to each residue in the set of 895 protein-protein interfaces. Information gain for a residue is calculated by summing up the information gain of its constituent atoms. Information gain of an atom is calculated from pair and single contact frequencies and using the square-bracketed term on the RHS of Eqn. 8.

# FIGURE CAPTIONS

**Fig. 1.** Schematic of the density of states for one probe macromolecule. A state is defined as the binding of a probe macromolecule to one of the many target macromolecules. Shaded area is the distribution of energies from non-specific binding of probe macromolecule to a decoy interface. $E_C$ is the energy of the typical "best" decoy interfaces to the probe interface. $E_N$ is the energy of the native complex.

**Fig. 2.** Schematic of lattice interface model. Two interfaces are aligned and pairwise energies calculated over all 25 sites. At each site sits one of 50 lattice "chemical" species or types. Black, white, and gray squares represent a few different species.

**Fig. 3.** The measured MI depends on the energy gap ratio $\varepsilon$ in a quadratic fashion for designed lattice interfaces. The energy gap ratio is defined as in Eqn. 7. Each point represents the MI and average energy gap of 1,000 designed probe-target pairs at one value of the selection strength $S$. Mutual information was measured in designed probe-target lattice interfaces via Eqn 1. Probe interfaces were paired with 10,000 random decoy interfaces to estimate $E_{AVG}$ and $E_C$ for calculating $\varepsilon$.

**Fig 4.** Spearman ranked correlation test between amino acid rankings obtained by information gain and conservation. 18 amino acids were ranked by information gain as calculated using Eqn. 8, and by conservation scores from Keskin et al (20). The correlation is 0.63 $(p < 0.003)$. Alanine and cysteine were excluded due to constraints of the original experiment cited in (20).

| Scheme | Mutual Information per contact (bits) | | |
|---|---|---|---|
| Residue | PDB | Shuffled | Corrected |
| I[1] | 0.0208 | 0.0065±0.0001 | 0.0143 |
| II[2] | 0.0119 | 0.0064±0.0002 | 0.0055 |
| III[3] | 0.0308 | 0.0093±0.0002 | 0.0215 |
| IV[4] | 0.0776 | 0.0222±0.0007 | 0.0554 |
| Atom | | | |
| Type23[5] | 0.1009 | 0.0040±0.0003 | 0.0969 |
| Type79[6] | 0.1885 | 0.0315±0.0004 | 0.1570 |

Table 1

| Residue | Information Gain (bits) |
|---------|-------------------------|
| ASP | 1.86 |
| TRP | 1.63 |
| LYS | 1.18 |
| GLU | 1.15 |
| PHE | 1.11 |
| ARG | 1.08 |
| LEU | 0.84 |
| CYS | 0.83 |
| TYR | 0.67 |
| HIS | 0.66 |
| ILE | 0.62 |
| ASN | 0.51 |
| PRO | 0.44 |
| MET | 0.44 |
| GLN | 0.40 |
| VAL | 0.33 |
| THR | 0.21 |
| SER | 0.20 |
| GLY | 0.08 |
| ALA | 0.04 |

Table 2

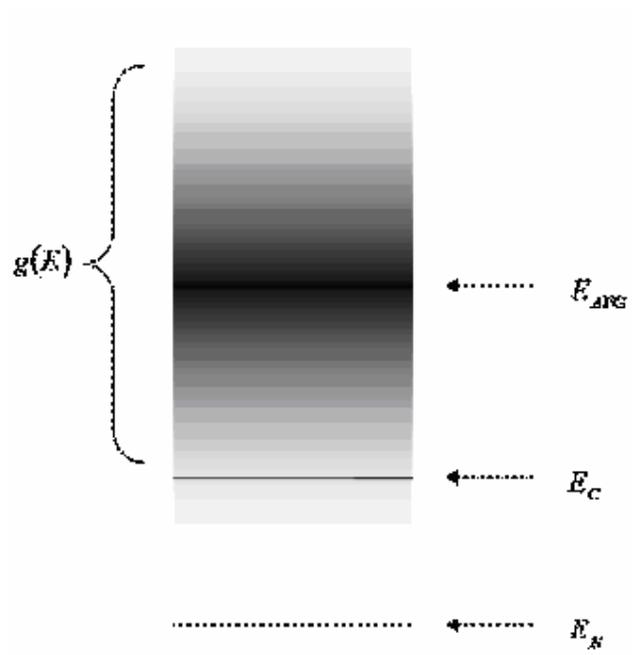

**Figure 1**

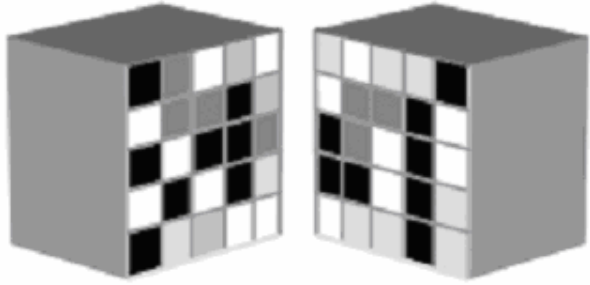

**Figure 2**

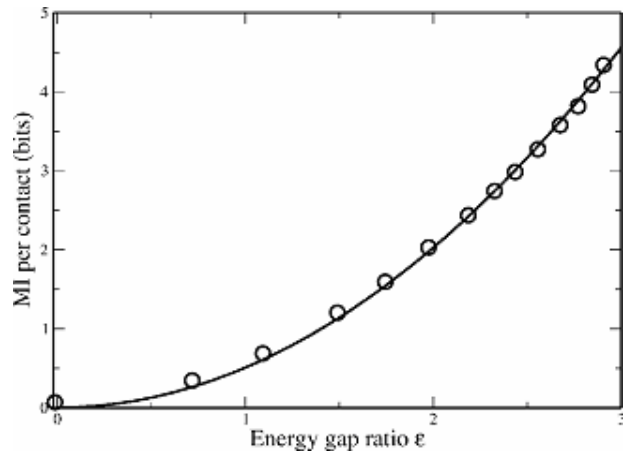

**Figure 3**

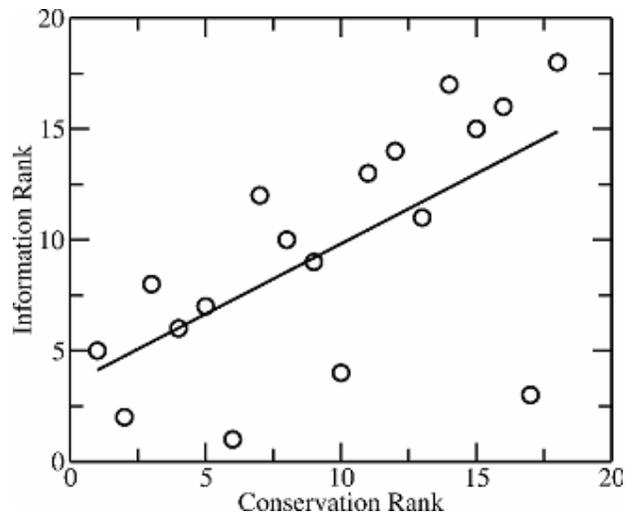

**Figure 4**

# Supplemental Materials

## Methods

### List of 895 protein-protein interfaces by PDB codes

Protein-protein interfaces were obtained from a previous work by Keskin et al (1). All 2-chain interfaces are indicated by a PDB code, followed by two letters which are the chain identifiers in the PDB file.

```
10gsAB 1a02FJ 1a07AC 1a0aAB 1a0hBC 1a0jCD 1a0mAB 1a0nAB 1a14HL
1a1uAC 1a2cHI 1a2lAB 1a2pBC 1a2yAB 1a37PB 1a38PB 1a3rHP 1a3rLP
1a3yAB 1a6aAC 1a6aBC 1a6uLH 1a7fAB 1a7vAB 1a8kAC 1a8mAB 1a8rAB
1a92AC 1a93AB 1aalAB 1aarAB 1aboAC 1ac6AB 1acyHP 1ae1AB 1ae9AB
1afa12 1afa13 1afrBD 1ag1OT 1agwAB 1ah8AB 1ahhAB 1ahwAC 1ahwEF
1ai0BD 1aikNC 1aipCE 1ajyAB 1ak4CD 1akeAB 1akjDE 1akmAB 1akmAC
1al04F 1al212 1al213 1al223 1antLI 1ao3AB 1ao7DE 1aohAB 1aoiAB
1aoiCD 1aonAO 1aplAC 1aq5AB 1aq5AC 1aqcAB 1aqdAC 1aqdBC 1ar814
1as4AB 1auvAB 1avoAC 1avoAM 1avoAN 1avpAB 1avwAB 1avzBC 1aw1AB
1ax4AD 1axcAC 1axcAE 1axdAB 1aym12 1aym13 1aym23 1azdAC 1azeAB
1azsBC 1azyAB 1azzAC 1b00AB 1b07AB 1b08AB 1b08AC 1b26AB 1b26AC
1b33BI 1b35BC 1b48AB 1b67AB 1b6bAB 1b77AB 1b77AC 1b8dAL 1b9bAB
1b9cAB 1b9lAB 1bazBD 1bb1AB 1bb1AC 1bb3AB 1bbrEF 1bccEJ 1bccFG
1bdtBD 1be3JK 1bev12 1bev13 1bev14 1bev23 1bfrAH 1bfrAW 1bgyBR
1bh8AB 1bho12 1bhq12 1bi4AC 1bj1HW 1bjjBC 1bjqAB 1bjrEI 1bogAC
1bogBC 1bonAB 1bqpAC 1br1AG 1br1AH 1brbEI 1brcEI 1brrBC 1brwAB
1bsxAX 1bt6AB 1bt6AC 1bt6BD 1btmAB 1bvrAB 1bvsAF 1bx2AC 1bx2BC
1bxkAB 1byfAB 1byoAB 1byuAB 1byzAB 1bzwAB 1bzxEI 1c08AC 1c08BC
1c09AC 1c14AB 1c17KM 1c17LM 1c1gAC 1c28AC 1c2rAB 1c2yAB 1c2yAE
1c41AB 1c41AE 1c4rBH 1c5fBC 1c6oAB 1c6vAX 1c72AB 1c78AB 1c8nAB
1c8oAB 1c94AB 1c9pAB 1c9tAG 1c9uAB 1ca0CD 1ca4AB 1ca4AC 1ca7AB
1ca7AC 1ca9AG 1cbiAB 1cblAB 1cd0AB 1cdtAB 1ce0AB 1ce0AC 1ce9AB
1cfsAC 1cfsBC 1cfyAB 1chkAB 1cjqAB 1cl7LH 1clxAD 1cm4DH 1cnt13
1cnt14 1cosAB 1cosAC 1cov12 1cov13 1cov14 1cpcAL 1cqxAB 1cseEI
1csgAB 1cu4HP 1cu4LP 1cudBC 1cunAB 1cunAC 1cvsAD 1cy9AB 1cydAB
1cydAD 1cz7AB 1cz7CD 1czpAB 1czvAB 1czyAD 1czzAD 1d0gAB 1d1mBA
1d2hAB 1d3bAB 1d3bAF 1d3bBC 1d3bFJ 1d5sAB 1d5wBC 1d7bAB 1d8hAC
1d9kAB 1d9kCP 1d9kDP 1d9uAB 1daoDF 1dbrAD 1dchAD 1dciAB 1dciAC
1de7HA 1debAB 1deeAD 1df1AB 1dfmAB 1dfnAB 1dg1GH 1dh3AC 1di0AB
1di0AE 1dipAB 1dirAB 1dj8CE 1dk7AB 1dkdAC 1dkdAE 1dkdBD 1dksAB
1dlgAB 1dlhAC 1dlhBC 1dowAB 1dpsAE 1dpsAH 1dpsAI 1dpsAK 1dpuAB
1ds5AF 1dt7AB 1dtjAB 1dubAB 1dubAC 1dubAD 1dvaHC 1dvaID 1dwuAB
1dx9CD 1dxxAD 1dylBD 1dz1AB 1dz4AB 1dzqAB 1e0bAB 1e2aAB 1e2aAC
1e3sAB 1e3sAC 1e4vAB 1e69AB 1e6jHP 1e79BG 1e79CG 1e7kAB 1e8aAB
1e92AB 1e92AC 1ea4DF 1eaiAC 1eawAB 1ebdAC 1eboAB 1eboAC 1ec5BC
1ecxAB 1ef1CD 1ef3AB 1ef7AB 1ef8AB 1ef8AC 1egjAL 1ehkBC 1ej3AB
1ej6BE 1ejbAB 1ejbAE 1ejmAB 1ejoHP 1ejoLP 1ek6AB 1ekxAB 1ekxAC
1em9AB 1emuAB 1eoiAB 1eoiAC 1eojAB 1epaAB 1eq2AB 1eq2AE 1eq8AB
1eq8AE 1esgAB 1et1AB 1etzHA 1eumAC 1evrFJ 1ew6AB 1ewaAB 1exzAC
1eygAC 1ez4AC 1ezsAD 1ezvEI 1ezvFG 1f05AB 1f08AB 1f1zAB 1f23AB
1f23AC 1f23DE 1f2dBD 1f2eAB 1f2iAH 1f2kAB 1f2lCD 1f2nAB 1f3jAP
1f3jBP 1f4kAB 1f4mAB 1f58LP 1f66CG 1f6fAC 1f6fBC 1f88AB 1f93EF
1f95AB 1f9fAD 1fakHI 1fb1AB 1fbiHX 1fbmAB 1fbmAE 1fbyAB 1fdqAB
1fduAC 1fe6AB 1fe6AD 1fe8BM 1fg9CE 1fguAB 1fi8AC 1fiuAC 1fj1AE
1fj1AF 1fj1BF 1fj1DE 1fjgGI 1flcAF 1fleEI 1flmAB 1fltVX 1fm6DE
1fmhAB 1fnsHA 1fntAB 1fntAd 1fntAG 1fntAH 1fntAI 1fntBc 1fntBC
1fntBI 1fntBJ 1fntCD 1fntCi 1fntCJ 1fntCK 1fntDE 1fntDh 1fntDL
```

```
1fntEF  1fntEg  1fntEL  1fntEM  1fntFf  1fntFG  1fntFM  1fntFN  1fntGe
1fntGH  1fntGN  1fntHI  1fntHN  1fntHV  1fntIa  1fntIb  1fntJa  1fntJK
1fntJZ  1fntKL  1fntKZ  1fntLM  1fntMN  1fo0AB  1fo0HB  1foeAD  1fosEF
1fpuAB  1fq3AB  1fqjAC  1fqjBE  1fs5AB  1fskAB  1fskFI  1fslAB  1ftaAD
1ftaBC  1fu1AB  1fuuAB  1fvfAB  1fvrAB  1fvvAD  1fx9AB  1fyhAE  1fytBD
1fzaAB  1fzaAC  1g0oAD  1g0uOP  1g0uOU  1g0uOV  1g1iAB  1g1kAB  1g2cAB
1g2cAC  1g2cAD  1g2cAE  1g2cBJ  1g2cDH  1g2yAB  1g39AB  1g39AD  1g39BD
1g3iGM  1g3iGR  1g4yBR  1g5gCF  1g5uAB  1g6rBH  1g6uAB  1g7kAB  1g83AB
1g8qAB  1g9iEI  1gaxAB  1gcoAB  1gcpBD  1gcqAC  1gcqBC  1gd2EF  1gd2HI
1gd8AG  1ge7AB  1gefAE  1gegAB  1gggAB  1ggiHP  1gh6AB  1gk4AB  1gk4AF
1gk4CE  1gk4DF  1gl1AI  1gl2AB  1gl2AC  1gl2AD  1gl2BC  1gl2BD  1gl2CD
1gl4AB  1gmjAB  1gmjBC  1gmjCD  1gmuBD  1gnwAB  1go4EF  1go4EH  1go4GH
1gp1AB  1gtuBD  1guyAC  1gwcBC  1gzuAB  1gzuAC  1h4xAB  1h59AB  1h5qAB
1h6kBZ  1h88AB  1h8oAB  1h97AB  1h9uBC  1hdcAC  1hdcAD  1hezAE  1hezCE
1hf9AB  1hfoAB  1hfoAC  1hg3AB  1hgdAD  1hgeAD  1hi6AC  1hiaBI  1himLP
1hiwAR  1hjaCI  1hjrAC  1hlcAB  1hleAB  1ho1AC  1hq3DF  1hqjAB  1hqjAC
1hqkAB  1hqkAE  1hqrBD  1hrdAB  1hrdAC  1hri12  1hri13  1hri14  1htmAB
1huuBC  1huxAB  1hv4BD  1hv8AB  1hvvAB  1hvvAC  1hvvAD  1hvvBC  1hvvBD
1hxyBD  1hy5AB  1hyhAB  1hyrAC  1hyrBC  1hz6BC  1hzdAB  1hzdAC  1hzdAD
1hzdCE  1hzzBC  1i01AB  1i01AD  1i0cAB  1i10AC  1i3rCE  1i4dAD  1i4k12
1i4kZ1  1i4lBD  1i4yAC  1i4yAG  1i55AB  1i59AB  1i5aAB  1i5kAD  1i5nAD
1i5nBC  1i7hAB  1i7hAC  1i7qAD  1i8fAB  1i8fAG  1i8kAC  1ia0AK  1iakAP
1iakBP  1iazAB  1ic2AB  1ic2BC  1ic2CD  1id2AC  1id3DF  1iesAB  1iesAC
1iesBF  1ifqAB  1igqAC  1ihjAC  1ihjBC  1ii2AB  1ijdAC  1ijxCD  1ik9AC
1ik9BC  1io6AB  1io7AB  1iodBG  1iqaAB  1iqdAC  1iqdBC  1irjAB  1irjBG
1iru12  1iruAG  1iruCK  1iruDK  1iruEM  1iruFG  1iruFM  1iruFN  1iruGH
1iruI1  1iruJ1  1iruJK  1iruKL  1iruKZ  1iruOP  1iruOU  1iruU2  1iruV2
1iruZ1  1irxAB  1is7AL  1it3AD  1ithAB  1j51AD  1j73CD  1j7dAB  1j8fBC
1j96AB  1j9kAB  1jb0IL  1jcjAB  1jd1AB  1jd1AF  1jd2HN  1jd2LO  1jd2MN
1jd2MO  1jd2NU  1jd2OU  1jd2OV  1jegAB  1jf7AB  1jfiAB  1jfiAC  1jfiBC
1jfmBC  1jfzAD  1jg3AB  1jgcAB  1jgcAC  1jgcBC  1jh5AB  1jhlHA  1jhlLA
1ji1AB  1ji5AC  1ji5BC  1jigBD  1jj2BT  1jj2R1  1jjoCE  1jk8AC  1jk8BC
1jl4BD  1jlvAB  1jm0DF  1jm7AB  1jmtAB  1jnhDH  1jnpAB  1jnrAD  1jplAE
1jpnAB  1jpsHT  1jpxDG  1jr3DE  1jrhLI  1js1XY  1js1XZ  1js4AB  1jstAD
1jthAB  1jthAD  1jthBD  1juqAE  1juqFC  1jwdAB  1jwhBD  1jxzAB  1jxzAC
1jy2NO  1jy2NP  1jy2OP  1jy2OS  1jy2PR  1jy2QR  1jy2QS  1jy6AB  1k05BC
1k1fAB  1k1fAC  1k1fDF  1k4cAC  1k4cBC  1k4dBC  1k4wAB  1k50AC  1k6jAB
1k7lCF  1k8kBG  1k8kFG  1k8rAB  1kb5BL  1kbaAB  1kcfAB  1kcgAC  1kcgBC
1kcrHP  1kcrLP  1kd8AB  1kepAB  1kf6DP  1kg0BC  1ki1BD  1kigHI  1kilAB
1kilAC  1kilAD  1kilAE  1kilBD  1kilBE  1kilCD  1kj4AP  1kjfAP  1kjhAP
1kkqAE  1klqAB  1kmeAB  1kq1AB  1kq1AH  1kqlAB  1kqsR1  1ksxAF  1kyoBR
1kyqAC  1l1oAC  1l1oCF  1l1oDF  1l2iAC  1l7cAC  1ldtTL  1lghAD  1lghGJ
1ljrAB  1lk3AL  1lldAB  1lmkAE  1lqvAC  1lt3DA  1lt3HA  1lvfAB  1mdyCD
1meyAC  1meyDF  1mj2AC  1mpgAB  1mr8AB  1ncaNL  1nrqHR  1nsiBC  1occCJ
1occDM  1om2AB  1otfAE  1otgAB  1otgAC  1pd212  1pma12  1pmaAB  1pmaAC
1pmaAH  1pmaAP  1pmaBY  1pmaBZ  1pmaIO  1pmaNO  1pmaZ1  1ppfEI  1psrAB
1qa9AB  1qbzAB  1qbzAC  1qd9AB  1qd9AC  1qeyAC  1qeyAD  1qghAE  1qghAH
1qghAI  1qghAK  1qjbAS  1qmoAB  1qu9AB  1qu9AC  1quqAD  1rbcSB  1rbdSB
1rbhSB  1rbiSB  1rhgAC  1rhgBC  1rrgAB  1rvf14  1rvv12  1rvvZ1  1rypEL
1rypOP  1rypOU  1rypRY  1sbwAI  1scjAB  1sebAC  1sebBC  1sebEG  1sebFG
1sfcBD  1sfcBE  1sfcBJ  1shkAB  1smvAB  1svfAB  1tafAB  1tawAB  1tecEI
1tf6BD  1tfxAC  1tgsZI  1tiiDC  1tiiEC  1tiiGC  1tiiHC  1tme23  1tvdAB
1unkAC  1wfaAB  1ydvAB  1zaaAC
```

# Supplemental Results

## Shuffling as a method of correcting biases in MI estimates

Here, we justify the subtraction of the shuffled MI from the measured MI by examining how the MI changes as a function of sample size.  For the case of measuring MI in contacts, the sample size is given by the total number of interface contacts in the set of PDBs of complexes.  We looked at the set of protein-protein complexes, which yields a total sample size of 156,566 contacts.  In conjunction with the 79 atom typing system, this gives a large enough sample size so that sampling is quite good (as revealed by the shuffled MI).  In this way, we can look at the behavior of MI both in the regime of good and poor sampling.  Mutual information, shuffled MI and corrected MI were measured for subsets of the PDB structures of complexes.  The number of structures in the subsets was increased gradually to include all 156,566 contacts (given by 895 interfaces).  The results are shown in SFig 1.  As the sample size increases, the shuffled MI approaches 0, and the measured MI approaches the corrected MI.  But even at low sample sizes (e.g. ~8,400 contacts), the corrected MI per contact has stabilized and is very close to the measured MI at high sample sizes (e.g. ~150,000 contacts).  At very low sampling, the corrected MI is expected to be quite unreliable: this shows up as an initially high corrected MI that subsequently and rapidly goes to the stable asymptotic corrected MI.  The stability of the corrected MI in conjunction with the decaying shuffled MI suggests that even when sampling is relatively poor, it is a good estimator of the true MI.

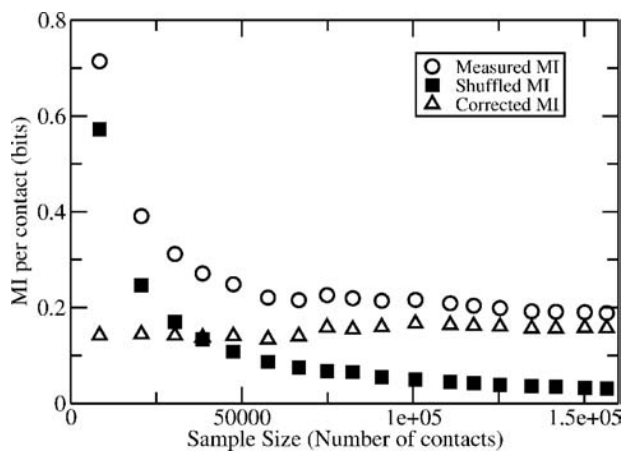

**SFig 1.**  MI per contact was measured in increasingly larger subsets of the complete set of protein-protein complexes.  The sizes of the subsets are given in terms of total number of interface contacts in each set.  MI per contact is shown as a function of the number of contacts.  Measured and shuffled MI begin at high values, highlighting the effect of small sample sizes and undersampling.  The measured MI (open circles) and the shuffled MI (closed squares) both decrease and approach asymptotic values as sample size increases.  The shuffled MI approaches 0 bits, while the measured MI converges to the corrected MI (open triangles).  The corrected MI is already quite stable after sample sizes of 8000.  Typing scheme used was the 79 atom-type.